\documentclass[conference,a4paper]{IEEEtran}

\usepackage{tcolorbox}
\usepackage{wrapfig}
\usepackage{float}
\usepackage[utf8]{inputenc} 
\usepackage[T1]{fontenc}
\usepackage{url}              
\usepackage{cite}             
\usepackage{bm}
\usepackage[cmex10]{amsmath}  
\usepackage{mleftright}       
\mleftright                   

\usepackage{graphicx}         
\usepackage{booktabs}         

\usepackage[utf8]{inputenc}
\usepackage[margin=0.5in]{geometry}
\usepackage{amsmath, amssymb, amsthm, nicefrac}
\usepackage{graphicx,subcaption}
\usepackage{tcolorbox}
\usepackage{enumerate} 
\usepackage{comment}
\usepackage{amsmath}
\usepackage[english]{babel}
\usepackage{amsthm}

\title{$t \geq 1$}

\newcommand{\linebreakand}{%
  \end{@IEEEauthorhalign}
  \hfill\mbox{}\par
  \mbox{}\hfill\begin{@IEEEauthorhalign}
}

\newtheorem{theorem}{Theorem}
\newtheorem{corollary}{Corollary}

\newtheorem{example}{Example}

\theoremstyle{remark}
\newtheorem{remark}{Remark}

\theoremstyle{definition}

\definecolor{cadmiumgreen}{rgb}{0.0, 0.42, 0.24}

\title{Coded Computing Meets Quantum Circuit Simulation: Coded Parallel Tensor Network Contraction Algorithm} 

\author{%
  \IEEEauthorblockN{%
    \begin{tabular}{ccc}
      Jin Lee && Sofía González-García \\
      University of California, Santa Barbara &&  University of California, Santa Barbara \\
        hojin@ucsb.edu && sofiagonzalezgarcia@ucsb.edu\\
        \vspace{2mm}\\
      Zheng Zhang && Haewon Jeong \\
      University of California, Santa Barbara &&
      University of California, Santa Barbara \\
        zhengzhang@ece.ucsb.edu &&
        haewon@ucsb.edu \\
    \end{tabular}
  }
}
\begin{document}
\maketitle
\begin{abstract}
Parallel tensor network contraction algorithms have emerged as the pivotal benchmarks for assessing the classical limits of computation, exemplified by Google's demonstration of quantum supremacy through random circuit sampling. However, the massive parallelization of the algorithm makes it vulnerable to computer node failures. In this work, we apply coded computing to a practical parallel tensor network contraction algorithm. To the best of our knowledge, this is the first attempt to code tensor network contractions. Inspired by matrix multiplication codes, we provide two coding schemes: 2-node code for practicality in quantum simulation and hyperedge code for generality. Our 2-node code successfully achieves significant gain for $f$-resilient number compared to naive replication, proportional to both the number of node failures and the dimension product of sliced indices. Our hyperedge code can cover tensor networks out of the scope of quantum, with degraded gain in the exchange of its generality.
\end{abstract}

\section{Introduction}
The search for quantum advantage hinges on the fact that the dynamics of quantum computers are hard to simulate with classical resources \cite{Bremner_2010, aaronson2010, Bremner_2016}, given the exponential growth of the state space with the number of qubits. Tensor networks are arguably the most important classical tool for simulating quantum circuits. Algorithms such as tensor network contractions \cite{markovshi, boixo, stepdependent,jet, liu_2024, huang2020classical, Pan_2022, kalachev2021classical, chenzhang, Villalonga_2019}, together with contraction order optimization \cite{Gray_2021} and parallelization by tensor index slicing \cite{chenzhang} lie at the heart of determining a beyond-classical threshold in quantum computer experiments. Google's quantum supremacy work relied on this classical benchmark \cite{arute2019quantum, morvan2023phase}. However, the scope of tensor network applications extends far beyond quantum circuit simulation. They are widely used as ans\"atze for many-body wavefunctions: Matrix Product States (MPS) for one-dimensional (1d) systems \cite{white_92, fannes1992finitely}, and their higher dimensional generalization in the form of Projected Entangled Pair States (PEPS) \cite{verstraete2004}. Moreover, tensor networks have found potential applications in machine learning \cite{stoudenmire2017supervised, Huggins_2019, mlexample1,mlexample3}, quantum chemistry \cite{chen2022using, MartiReiher, Lee_2023} and other optimization problems \cite{Kourtis_2019, Liu_2023}.

Despite all the great success and extensive recognition however, tensor network simulation faces reliability challenges due to the massive parallelization for efficiency. To effectively simulate state-of-the-art quantum circuits, tensor network contraction algorithms must be distributed across millions of classical machines.  Distributing workloads to billions of threads inevitably increases the rate of errors. At the same time, as we are incorporating more diverse devices with different computational capacities (e.g., computation-specific accelerators, quantum processors), the variance in node response time significantly increases. Managing such uncertainties entirely at the operating systems (OS) level puts too much strain on the already overloaded OS. 

In this paper, we propose a node failure tolerant scheme applied at the algorithm level by applying coded computing to tensor network contraction. Coded computing, also known as algorithm-based fault tolerance (ABFT), has proposed innovative ways to add redundancy in the computation using error-correcting codes. Coded computing strategies were developed for essential numerical algorithms such as matrix multiplication~\cite{matdot,polycode,wang2018coded,severinson2018block,mallick2018rateless,jeong2021e}, FFT~\cite{jeong2018masterless,yu2017coded}, and matrix factorization~\cite{quangQR,nguyen2023coded}. Matrix multiplication, in particular, has been a focal point of extensive research as it is an important backbone of scientific and machine learning applications. 

Our contributions of this paper are for both tensor network and coded computing community. For the tensor network community, we provide a first work on  parallel tensor network contraction algorithm that can be resilient to stragglers and failures. For the coded computing community, we provide elegant generalization of matrix multiplication codes, to the extremely expressive high-dimensional linear algebraic framework of tensor network.

\section{Background and Notations}
In this section, we provide a concise overview of tensor network contraction and its parallel algorithms, as well as the notations employed throughout the paper. For a more extensive review of this subject, refer to \cite{tnguide,nutshell,quantummanybodyex2,quantummanybodyex4}. 

\subsection{Tensor Networks and contraction}

\begin{figure}[t]
    \centering
    \includegraphics[width=0.35\textwidth]{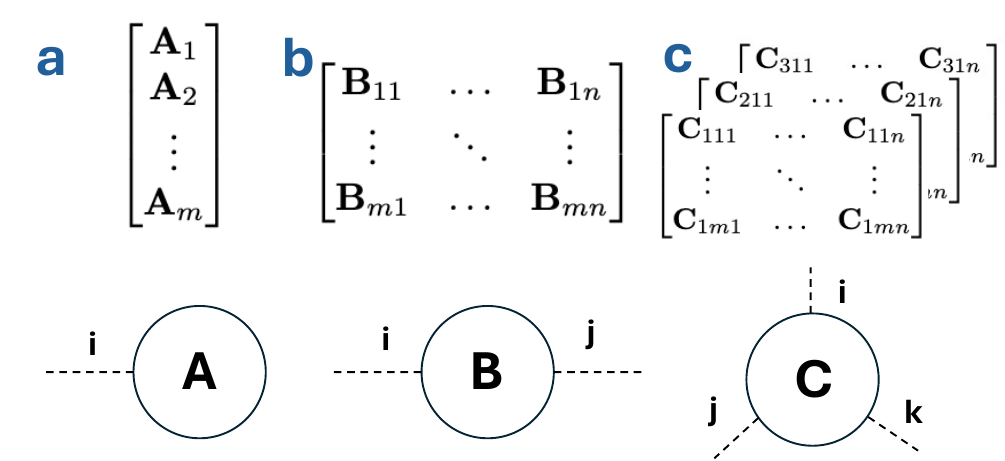}
    \caption{\small{Diagrammatic representation and corresponding array for ($\boldsymbol{a}$) one dimensional tensor $\boldsymbol{A}_{i}$ ($\boldsymbol{b}$) two dimensional tensor $\boldsymbol{B}_{i,j}$ ($\boldsymbol{c}$) three dimensional tensor $\boldsymbol{C}_{i,j,k}$}}
    \label{fig:tensor_ex}
    \vspace{-7pt}
\end{figure}

Tensors are multidimensional arrays that extend the concepts of one-dimensional vectors and two-dimensional matrices to higher dimensions, and the tensor network formalism represents linear operations between tensors using graph structures. A tensor network consists of nodes and edges, where each node represents a tensor, and each edge connected to the node represents the index of the tensor. Diagrammatic representations of one-dimensional tensor $\boldsymbol{A}_{i}$, two-dimensional tensor $\boldsymbol{B}_{i,j}$, and three-dimensional tensor $\boldsymbol{C}_{i,j,k}$ are shown in Fig \ref{fig:tensor_ex}. We use bold text with its index labels given in subscript for tensors. E.g., $\boldsymbol{B}_{i,j}$ is a two-dimensional tensor with indices labeled as $i$ and $j$; note that this is not a scalar entry $B[i,j]$. $\boldsymbol{C}_{i,j,k}$ is a three-dimensional tensor and $\boldsymbol{C}_{1,j,k}$ is a two-dimensional tensor when the first index is fixed to  $i=1$. 

\begin{figure}[t]
    \centering
    \includegraphics[width=0.5\textwidth]{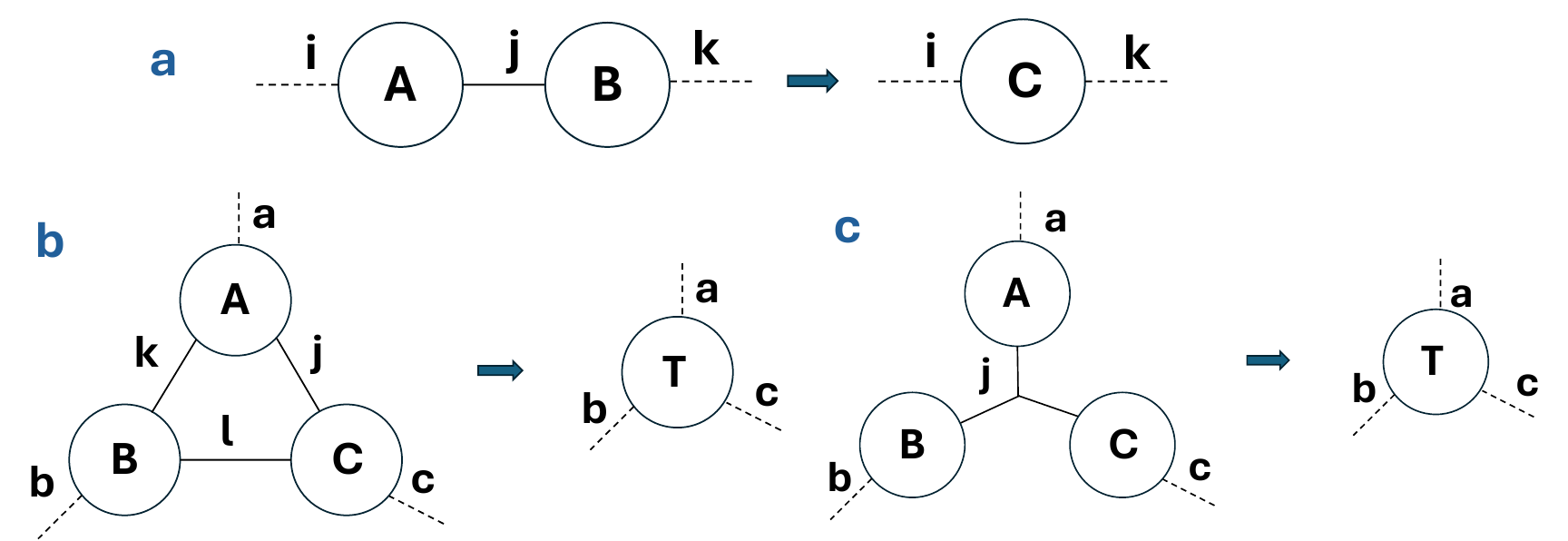}
    \caption{\small{Diagrammatic representation of tensor network contraction for ($\boldsymbol{a}$) $\sum_{j}\boldsymbol{A}_{i,j}\boldsymbol{B}_{j,k}$   ($\boldsymbol{b}$) $\sum_{k,j,l}\boldsymbol{A}_{a,j,k}\boldsymbol{B}_{b,k,l}\boldsymbol{C}_{c,l,j}$  
 ($\boldsymbol{c}$) $\sum_{j}\boldsymbol{A}_{j,a}\boldsymbol{B}_{b,j}\boldsymbol{C}_{j,c}$}}
    \label{fig:contraction_ex}
    \vspace{-7pt}
\end{figure}

An edge connecting two nodes in a tensor network represents a \emph{contraction} operation, which generalizes an inner product of the dimension---pairwise multiplication followed by summation. For instance, matrix multiplication can be represented as Fig.~\ref{fig:contraction_ex}(a), where two nodes $\boldsymbol{A}$ and $\boldsymbol{B}$ are connected through the edge $j$, representing $C[i,k] = \sum_j A[i,j] B[j,k]$. With a slight abuse of notations, we will also write it in a tensor form as $\boldsymbol{C}_{i,k} = \sum_j \boldsymbol{A}_{i,j} \boldsymbol{B}_{j,k}$. In Fig.~\ref{fig:contraction_ex}(b), we illustrate contracting three edges (indices $j,k,l$) between three tensors $\boldsymbol{A, B, C}$, resulting in a tensor $\boldsymbol{T}_{a,b,c} = \sum_{j,k,l} \boldsymbol{A}_{a,j,k}\boldsymbol{B}_{b,k,l} \boldsymbol{C}_{c,l,j}$. The edges $j,k,l$ are called \emph{closed edges} and $a,b,c$ are called \emph{open edges.}  Note that the indices of closed edges disappear after the contraction and the final tensors are indexed only by open edges. With another slight abuse of notation, we use the terms edge and index interchangeably in this paper. In tensor networks, there can also be \emph{hyperedges} that connect more than two nodes as shown in Fig.~\ref{fig:contraction_ex}(c). We call an index shared by $m$ tensors as an $m$-node index. Based on this diagrammatic representation, we refer an index shared by more than two tensors as a hyperedge. 

When there are multiple closed edges like in Fig.~\ref{fig:contraction_ex}(b), contraction is done sequentially, and the order of contraction does not affect the final outcome. However, the order of contraction can significantly change the computational complexity, and choosing an optimal contraction order is a NP-complete problem\cite{markovshi}. Markov and Shi \cite{markovshi} proved that the computational complexity of the optimal contraction order for a tensor network is decided by its graph theoretical property. Hence, the topology of a tensor network decides its computational cost.

\subsection{Parallel tensor network contraction} \label{sec:parallel}
In a nutshell, a tensor network is an extremely expressive formalism that represents a series of arbitrary product sum operations for high-dimensional tensors. As the tensor network becomes more complex and the sizes of tensors become large, tensor network contraction can be very computationally expensive. To scale up the computation, researchers have proposed parallel tensor network algorithms~\cite{chenzhang} and showed breakthroughs in quantum circuit simulation\cite{indexslicing,stepdependent,jet}.



\begin{figure}[t]
    \centering
    \includegraphics[width=0.5\textwidth]{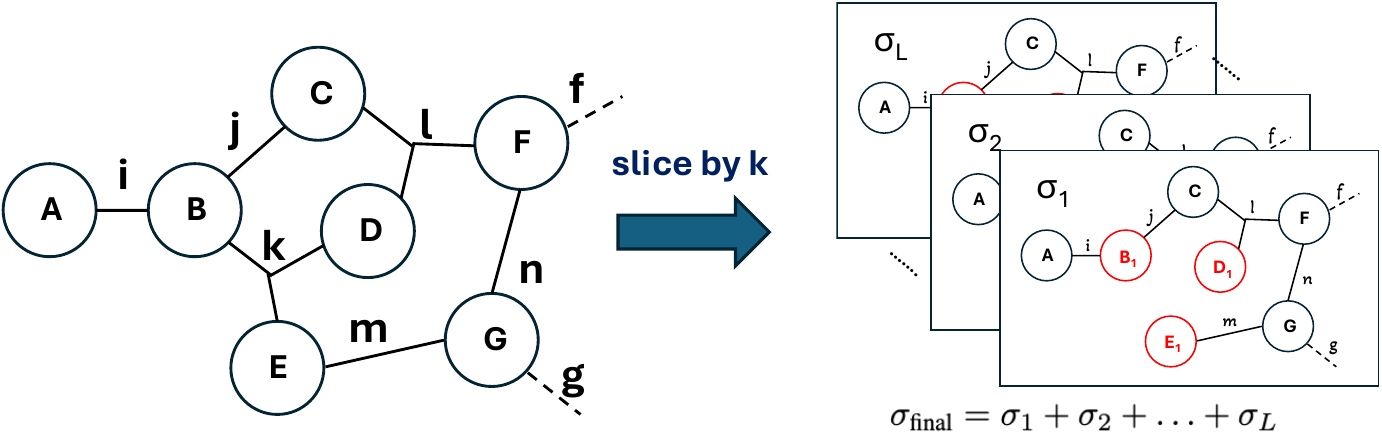}
    \caption{\small{Parallelization scheme by slicing index $k$ for the sample tensor network given in \ref{sec:parallel}}.}
    \label{fig:parallel}
    \vspace{-7pt}
\end{figure}

In this paper, we build up on the work by Chen and Zhang~\cite{chenzhang}, that decomposed 
tensor network contraction into an highly parallel algorithm. To explain the parallelization scheme, we will use an example network given in Figure~\ref{fig:parallel}: 
\begin{align}
    \boldsymbol{T}_{f,g} &= \sum_{i,j,k,l,m,n}\boldsymbol{A}_{i}\boldsymbol{B}_{i,j,k}\boldsymbol{C}_{j,l}\boldsymbol{D}_{k,l}\boldsymbol{E}_{k,m}\boldsymbol{F}_{l,n,f}\textbf{G}_{m,n,g} \nonumber \\ 
    &=\sum_{i,j,l,m,n}(\boldsymbol{A}_{i}\boldsymbol{B}_{i,j,1}\boldsymbol{C}_{j,l}\boldsymbol{D}_{1,l}\boldsymbol{E}_{1,m}\boldsymbol{F}_{l,n,f}\textbf{G}_{m,n,g}) \nonumber \\ 
    &+\sum_{i,j,l,m,n}(\boldsymbol{A}_{i}\boldsymbol{B}_{i,j,2}\boldsymbol{C}_{j,l}\boldsymbol{D}_{2,l}\boldsymbol{E}_{2,m}\boldsymbol{F}_{l,n,f}\textbf{G}_{m,n,g}) \nonumber \\ 
    &\quad \quad \qquad \quad \quad \qquad \quad \quad \qquad \vdots \nonumber \\ 
    &+\sum_{i,j,l,m,n}(\boldsymbol{A}_{i}\boldsymbol{B}_{i,j,L}\boldsymbol{C}_{j,l}\boldsymbol{D}_{L,l}\boldsymbol{E}_{L,m}\boldsymbol{F}_{l,n,f}\textbf{G}_{m,n,g})  \nonumber\\ 
    &=\boldsymbol{\sigma}_{1}+\boldsymbol{\sigma}_{2}+\ldots+\boldsymbol{\sigma}_{L} = \boldsymbol{\sigma}_\text{final},
\end{align}
where $\boldsymbol{\sigma}_{\tilde{k}}$ for $\tilde{k} \in \{ 1,2,\cdots, L\}$ is given as:
\begin{equation}\label{eq:sigma_k}
    \boldsymbol{\sigma}_{\tilde{k}} = \sum_{i,j,l,m,n}\boldsymbol{A}_{i}\boldsymbol{B}_{i,j,\tilde{k}}\boldsymbol{C}_{j,l}\boldsymbol{D}_{\tilde{k},l}\boldsymbol{E}_{\tilde{k},m}\boldsymbol{F}_{l,n,f}\textbf{G}_{m,n,g}.
\end{equation}
We say that the tensor network is \emph{sliced}
by the index $k$, and each $\boldsymbol{\sigma}_{\tilde{k}}$ is \emph{a sliced partition} of the original tensor network. 
Chen and Zhang~\cite{chenzhang} proposed an algorithm that distributed the computation of each sliced partition to a separate compute node and summing them up at the end to obtain $\boldsymbol{\sigma}_{\text{final}}$. For computing the sliced partition  $\boldsymbol{\sigma}_{\tilde{k}}$, no inter-node communication is required, which makes the algorithm embarrassingly parallel. In Figure~\ref{fig:parallel}, we show the diagrammatic representation of slicing: the corresponding edge for the slicing index $k$ is erased and the connected tensors $\boldsymbol{B}_{i,j,k}$, $\boldsymbol{D}_{k,l}$, $\boldsymbol{E}_{k,m}$ are replaced with sliced subtensors (e.g., $\boldsymbol{B}_{i,j,1}$, $\boldsymbol{D}_{1,l}$, $\boldsymbol{E}_{1,m}$)).

For the rest of the paper ,we further simplify the notations by abstracting out all parts not related to the slicing operation as  $\boldsymbol{T}_{-}$, where $`-'$ denotes all indices other than the slicing index. E.g., \eqref{eq:sigma_k} can be written as:
\begin{equation}
    \boldsymbol{\sigma}_{\tilde{k}} = \sum_{-}(\boldsymbol{B}_{\tilde{k},-}\boldsymbol{D}_{\tilde{k},-}\boldsymbol{E}_{\tilde{k},-})\boldsymbol{T}_{-}.
\end{equation}
Tensors $\boldsymbol{A,C,F,G}$ that are not connected to the edge $k$ are all abstracted into a big tensor $\boldsymbol{T}_{-}$. Finally, we say two edges are \emph{adjacent} if they are connected with the same node and \emph{non-adjacent} if they do not share a node. 

\section{System Model and Problem Statement}
\label{sec:system}

\begin{figure*}[t]
    \centering
    \includegraphics[width=0.85\textwidth]{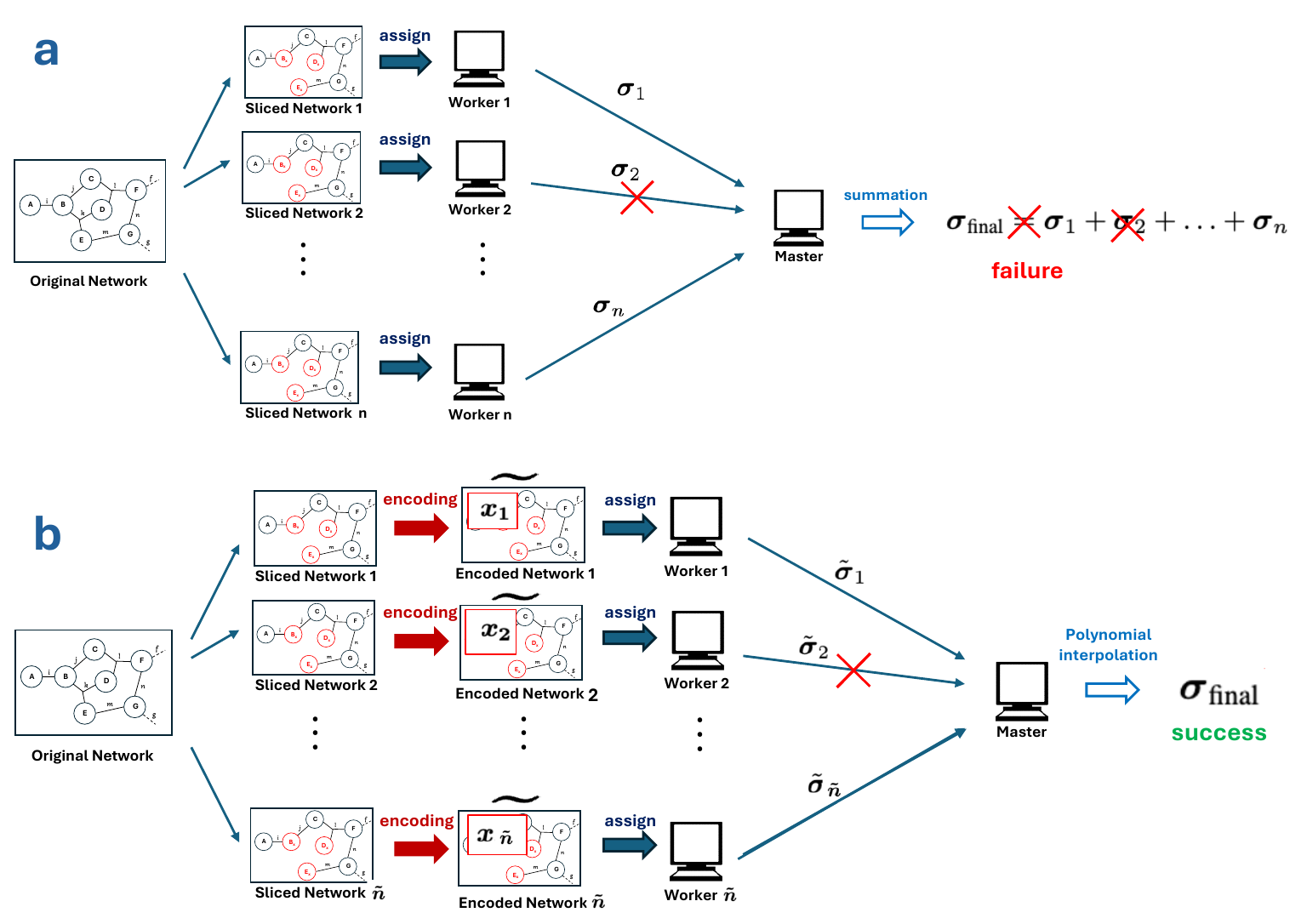}
    \caption{\small{($\boldsymbol{a}$) System model for distributed parallel tensor network contraction without coding. Failure on any computational node will result as a failure of entire system ($\boldsymbol{b}$) System model for distributed coded parallel tensor network contraction. Even with some node failures, as long as the number of successful computational nodes is over $f$-resilient number, desired final outcome will always be retrieved.}}
    \label{fig:system}
    \vspace{-7pt}
\end{figure*}

The goal of this paper is to develop coded computing strategies for parallel tensor network contraction for a given tensor network $G(V,E)$. For parallel tensor network contraction, we adopt Chen \& Zhang's algorithm~\cite{chenzhang} described in Section~\ref{sec:parallel}.We assume a computing model similar to \cite{matdot,polycode} which consists of a master compute node (\emph{master}) that distributes computation inputs and assign tasks to  worker compute nodes (\emph{workers}). Workers then perform the assigned computation job and report the result back to the master. With Chen \& Zhang's algorithm\cite{chenzhang}, each worker will compute a sliced tensor network $\boldsymbol{\sigma}_k$ and the master node will aggregate the outputs to obtain $\boldsymbol{\sigma}_\text{final} = \sum_k \boldsymbol{\sigma}_k$ as shown in Fig.~\ref{fig:system}.

An important distinction from the model given in \cite{matdot} is that for high-dimensional tensor contraction, it is often compute-bound rather than storage-bound. 
Thus, instead of a storage constraint, we impose that each worker in the coded computing scheme cannot have computational costs exceeding the original uncoded algorithm, i.e., when a worker performs coded computation on encoded tensor network  $\tilde{\boldsymbol{\sigma}}_k$, it should satisfy the following: 
\vspace{-0.3em}
\begin{equation}\label{eq:comp_bound}
    \text{comp. complexity}(\tilde{\boldsymbol{\sigma}}_k) \leq \text{comp. complexity}(\boldsymbol{\sigma}_k).
\end{equation}
Computation and memory requirements in tensor network contraction is fully determined by the network topology\footnote{This is assuming that the algorithm does not optimize for any sparsity in the tensor data.}, and thus having the same topology with same tensor dimensions implies the same computation cost.  

We assume that in the uncoded algorithm, the tensor network $G$ is sliced multiple times at $n$ edges, each of which is a $m_{i}$-node edge with its index dimension $L_{i}$ ($i={1,2,\ldots,n}$). This produces $N=\prod_{i=1}^{n}L_{i}$ sliced tensor networks and each of them will be distributed to individual workers. After all $N$ workers complete their computation, the master aggregates and obtains $\boldsymbol{\sigma}_\text{final}$. We further make an assumption that all sliced edges are \emph{non-adjacent} to each other. 

We assume that there can be $f$ worker failures and we cannot recover any partial result from the failed worker. We define \emph{$f$-resilient number} as the number of total workers required in the system to retrieve $\boldsymbol{\sigma}_\text{final}$ in case of arbitrary $f$  node failures. We will compare our coding strategy with a naive replication strategy, which has a $f$-resilient number of $N(f+1)$, as we need at least $(f+1)$ replicas of each worker. We now define our problem below. 

\textbf{Problem Statement.}
For a tensor network contraction parallelized by
slicing $n$ non-adjacent indices, each of which is $m_{i}$-node and $L_{i}$-dimensional ($i={1,2,\ldots,n}$), our aim is to design a coded computing strategy that can be resilient to any $f$ failures while maintaining the same computational complexity per worker and achieve a $f$-resilient number smaller than $N(f+1)$ where $N=\prod_{i=1}^{n}L_{i}$.

\begin{remark}
The assumption that all sliced indices are non-adjacent to each other makes our problem more tractable and has practical justification. In \cite{stepdependent}, the authors suggested a step-dependent iterative heuristics to find indices to slice by choosing an index that reduces the computational complexity the most (approximately) at each iteration. With this algorithm, indices connected to tensors already sliced in previous steps are not likely to be chosen.
Nonetheless, developing a coding scheme for tensor networks allowing adjacent sliced indices is an interesting open question.    
\end{remark}

\section{Main Results}
Coded computing strategies we propose in this paper are based on polynomials, similar to polynomial codes\cite{polycode} and MatDot/PolyDot codes\cite{matdot} for distributed matrix multiplication. 
%
I.e., inputs are encoded with polynomials, and after a set of multilinear computations, output will be encoded in a polynomial of a higher degree. The $f$-resilient number of the code is determined  by the degree of the resulting polynomial. If the resulting output polynomial is of $d$-th order, then with any $(d+1)$ successful workers, all its coefficients can be retrieved by an efficient polynomial interpolation algorithm \cite{polyinterpolation,polycode}. For $f$ worker failures, the $f$-resilient number is $(d+f+1)$, and the gain compared to naive replication is 
\begin{equation}\label{eq:gain}
    \Delta=N(f+1)-(d+f+1)=(N-1)(f+1)-d,
\end{equation}
where $N=\prod_{i=1}^{n}L_{i}$.
Hence, the goal of code design is to minimize the output polynomial degree while ensuring that all terms relevant to the tensor network contraction are retrievable from the polynomial coefficients.

In the following subsections, we provide two coding schemes and their $f$-resilient numbers. The first code is for tensor networks commonly used in quantum simulation, which are the main interest of contemporary tensor network studies. The second code is for general tensor networks, applicable for any tensor network as long as sliced indices are non-adjacent to each other, which is also a practical assumption.

\subsection{2-node Code for Quantum Simulation}

While $m$-node hyperedges can represent general tensor operations, in many practical applications, most edges are conventional 2-node edges. For instance, tensor networks for quantum simulation often follow the Projected Entangled-Pair States (PEPS) structure\cite{verstraete2004}, which form a 2D-grid graph that require only 2-node edges. PEPS is also commonly used in condensed matter physics~\cite{Corboz_2014a, Corboz_2014b, corboz_2017, Zheng_2017, Ponsioen_2019, Chen_2020}. We thus first develop a coding scheme specifically designed for slicing 2-node indices, which we refer to as the \emph{2-node code}.

\begin{theorem} \label{thm:2-node}
    For a tensor network contraction parallelized by
slicing at $n$ non-adjacent indices, each of which is $2$-node and $L_{i}$-dimensional ($i={1,2,\ldots,n}$), $f$-resilient number of following is achievable:
\begin{equation}\label{eq:threshold2}
f+2\prod_{i=1}^{n}L_{i}-1.
\end{equation}
The gain compared to replication for $f$ failures is:
\begin{equation}\label{eq:gain2}
\Delta=(\prod_{i=1}^{n}L_{i}-1)(f-1).
\end{equation}
\end{theorem}

\begin{proof}
We define template polynomials for $i=1,\cdots, n$:
\begin{align*}
    p_i^{(1)}(x) & =1+x+x^2+\ldots+x^{L_i-1}, \\ 
    p_i^{(2)}(x)&=x^{L_i-1}+x^{L_i-2}+x^{L_i-3}+\ldots+1,
\end{align*}
and then encoding polynomials for index $i$ are $p_i^{(1)}(x^{\prod_{k=1}^{i-1}L_{k}})$ with each coefficient substituted with the sliced tensors. E.g., for $i=2$, let us call the two tensors connected to this index as $\mathbf{B}^{(1)}$ and $\mathbf{B}^{(2)}$. Then, the encoded tensor polynomials are:
\begin{align*}
    \Tilde{\boldsymbol{B}}^{(1)}_{-}(x) &= \mathbf{B}^{(1)}_{1,-} + \mathbf{B}^{(1)}_{2,-} x^{L_1} + \cdots + \mathbf{B}^{(1)}_{L_2,-} x^{L_1 (L_2-1)}, \\
     \Tilde{\boldsymbol{B}}^{(2)}_{-}(x) &= \mathbf{B}^{(2)}_{1,-} x^{L_1 (L_2-1)} + \mathbf{B}^{(2)}_{2,-} x^{L_1 (L_2-2)} + \cdots + \mathbf{B}^{(2)}_{L_2,-}. 
\end{align*}
Note that all of the desired terms of the product $\Tilde{\boldsymbol{B}}^{(1)}_{-}(x)\Tilde{\boldsymbol{B}}^{(2)}_{-}(x)$ are aligned in the coefficient of  $x^{L_1 (L_2-1)}$. 

Likewise, the product of encoded tensors of index $i$ is a polynomial of $2(L_i-1)\prod_{k=1}^{i-1}L_k$ order which all desired terms are confined in $(L_i-1)\prod_{k=1}^{i-1}L_k$ order term. Hence, product of all encoded tensors is a polynomial of $2((\prod_{i=1}^nL_i) - 1)$ order which all desired terms are confined in $((\prod_{i=1}^nL_i) - 1)$ order term. With any successful $(2(\prod_{i=1}^nL_i)-2+1)$ encoded tensor networks each using different arbitrary constant $x$, coefficient of $((\prod_{i=1}^nL_i) - 1)$ order term can be retrieved, which is the final outcome of the original tensor network. Please refer to Appendix A for further details on the proof. \end{proof} 

Additionally, PEPS-structure tensor networks typically  consist of uniform dimensions for all closed edges. Hence, we provide a following corollary for this special case.

\begin{corollary}
When  $m_i=2, L_{i}=L > 2$ for all $i$, the following $f$-resilient number is achievable:
\begin{equation}
f+2L^{n}-1. 
\end{equation}
The gain compared to replication for $f$ failures is:
\begin{equation}
\Delta=(L^n-1)(f-1).  
\end{equation}
\end{corollary}

\begin{example}[2-node code]
We provide an example for 2-node code given in Theorem~\ref{thm:2-node} when we have two sliced edges with $m_1=2, L_1=4$ and $m_2=2, L_2=3$. The contraction equation can be written as:
\begin{equation}
    \boldsymbol{\sigma}_{a,b}=\sum_{-}(\boldsymbol{A}^{(1)}_{a,-}\boldsymbol{A}^{(2)}_{a,-}\boldsymbol{B}^{(1)}_{b,-}\boldsymbol{B}^{(2)}_{b,-})\boldsymbol{T}_{-},
\end{equation}
where $a\in\{1,2,3,4\}$ and $b\in\{1,2,3\}$ represent sliced indices. We encode the sliced tensors as follows: 
\begin{align*}
    \Tilde{\boldsymbol{A}}^{(1)}_{-} &=\boldsymbol{A}^{(1)}_{1,-} + \boldsymbol{A}^{(1)}_{2,-}x + \boldsymbol{A}^{(1)}_{3,-}x^{2}+\boldsymbol{A}^{(1)}_{4,-}x^{3}, \\ 
    \Tilde{\boldsymbol{A}}^{(2)}_{-} &=\boldsymbol{A}^{(2)}_{1,-}x^3 + \boldsymbol{A}^{(2)}_{2,-}x^2 + \boldsymbol{A}^{(2)}_{3,-}x+\boldsymbol{A}^{(2)}_{4,-}, \\ 
\Tilde{\boldsymbol{B}}^{(1)}_{-} &=\boldsymbol{B}^{(1)}_{1,-} + \boldsymbol{B}^{(1)}_{2,-}x^4 + \boldsymbol{B}^{(1)}_{3,-}(x^4)^2, \\
\Tilde{\boldsymbol{B}}^{(2)}_{-}&=\boldsymbol{B}^{(2)}_{1,-}(x^4)^2 + \boldsymbol{B}^{(2)}_{2,-}x^4 + \boldsymbol{B}^{(2)}_{3,-},
\end{align*}
where $x$ can be substituted with a distinct evaluation point at each worker. After the local contraction operation, the result at each worker is: 
\begin{align*}
    \Tilde{\boldsymbol{\sigma}} &=\sum_{-}(\Tilde{\boldsymbol{A}}^{(1)}_{-}\Tilde{\boldsymbol{A}}^{(2)}_{-}\Tilde{\boldsymbol{B}}^{(1)}_{-}\Tilde{\boldsymbol{B}}^{(2)}_{-})\boldsymbol{T}_{-} \\ 
    &= (\boldsymbol{\sigma}_{1,1}+\boldsymbol{\sigma}_{2,1}+\boldsymbol{\sigma}_{3,1}+...+\boldsymbol{\sigma}_{4,3})x^{11}+ \ldots \\ 
    &=\boldsymbol{\sigma}_{\text{final}}x^{11}+ \ldots =\text{poly}(x^{22}).
\end{align*}
Hence, with any $(22+1)$ successful workers with distinct evaluation points, we can perform polynomial interpolation to retrieve $\boldsymbol{\sigma}_{\text{final}}$. The gain compared to naive replication is
\begin{equation*}
   \Delta=12(f+1)-(22+f+1)=11(f-1) \hspace{4pt}(\text{workers}), 
\end{equation*}
which is consistent with Theorem~\ref{thm:2-node}.
\end{example}

\subsection{Hyperedge Code for General Case}

The tensor network framework is increasingly gaining recognition across various communities outside of the quantum field, and tensor network structures that incorporate hyperedges may find more use in the future. We thus provide a more general coding scheme beyond the 2-node code, suitable for any tensor network, which we refer to as the \emph{hyperedge code}.  We establish an achievable scheme for hyperedge codes in Theorem~\ref{thm:hyperedge}. The provided code works for any tensor network as long as two sliced edges are not adjacent to each other.

\begin{theorem} \label{thm:hyperedge}
    For a tensor network contraction parallelized by
slicing at $n$ non-adjacent indices, each of which is $m_{i}$-node and $L_{i}$-dimensional ($i={1,2,\ldots,n}$), $f$-resilient number of following is achievable: 
\begin{equation}\label{eq:threshold1}
f+\prod_{i=1}^{n}\frac{(m^{L_{i}}_{i}-1)}{(m_{i}-1)}.
\end{equation}

The corresponding gain compared to naive replication is
\begin{equation}\label{eq:gain1}
\Delta=(\prod_{i=1}^{n}L_{i}-1)(f+1)-\prod_{i=1}^{n}\frac{(m^{L_{i}}_{i}-1)}{(m_{i}-1)}+1.
\end{equation}
\end{theorem}

\begin{proof} We define template polynomial for $i=1,\cdots, n$:
\begin{align*}
    p_{i}(x)=1+x+x^{1+m_i}+x^{1+m_i+m_i^2}+\ldots+x^{1+\dots+m_i^{L_i-2}},
\end{align*}
and then the encoding polynomial for index $i$ is $p_i(x^{\prod_{j=1}^{i-1}\frac{m_j^{L_j}-1}{m_j-1}})$ with each coefficient substituted with the sliced tensors. E.g., for $i=2$, let us call the $m_2$ tensors connected to this index as $\mathbf{B}^{(j)}$ where $j=1,\ldots,m_2$. Then, the encoded tensor polynomials are:
\begin{align*}\Tilde{\boldsymbol{B}}^{(j)}_{-}(x) = &\mathbf{B}^{(j)}_{1,-} + \mathbf{B}^{(j)}_{2,-} x^{\frac{m_1^{L_1}-1}{m_1-1}} + \cdots \\
    &+ \mathbf{B}^{(j)}_{L_2,-} (x^{\frac{m_1^{L_1}-1}{m_1-1}})^{1+\ldots+m_2^{L_2-2}},
\end{align*}
Note that every desired term of the product $\prod_{j=1}^{m_2}\Tilde{\boldsymbol{B}}^{(j)}_{-}(x)$ is each confined in different order term only by itself, and the resulting polynomial is $\frac{m_1^{L_1}-1}{m_1-1}\cdot\frac{m_2^{L_2}-m_2}{m_2-1}$ order.
Likewise, the product of encoded tensors of index $i$ is a polynomial of $\frac{m_1^{L_1}-1}{m_1-1}\cdot\frac{m_2^{L_2}-1}{m_2-1}\cdot\frac{m_3^{L_3}-1}{m_3-1}\cdot...\cdot\frac{m_i^{L_i}-m_i}{m_i-1}$ order which each desired term is confined in different order term only by itself. Hence, product of all encoded tensors is a polynomial of $(\prod_{j=1}^{n}\frac{m_j^{L_j}-1}{m_j-1})-1$ order which each desired term is confined in different order term only by itself. With any successful $\prod_{j=1}^{n}\frac{m_j^{L_j}-1}{m_j-1}$ encoded tensor networks each using different arbitrary constant $x$, all the desired outcomes of sliced tensor networks can be retrieved and the final outcome of the original tensor network can be acquired by their summation. Please refer to Appendix B for further details on the proof.\end{proof} 

\begin{example} [Hyperedge code]
We provide an example for hyperedge code given in Theorem~\ref{thm:hyperedge} when we have two sliced edges with $m_1=3, L_1=2$ and $m_2=4, L_2=2$. The contraction equation can be written as:
\begin{equation*}\boldsymbol{\sigma}_{a,b}=\sum_{-}(\boldsymbol{A}^{(1)}_{a,-}\boldsymbol{A}^{(2)}_{a,-}\boldsymbol{A}^{(3)}_{a,-}\boldsymbol{B}^{(1)}_{b,-}\boldsymbol{B}^{(2)}_{b,-}\boldsymbol{B}^{(3)}_{b,-}\boldsymbol{B}^{(4)}_{b,-})\boldsymbol{T}_{-},
\end{equation*}
\vspace{-0.3em}
where $a\in\{1,2\}$ and $b\in\{1,2\}$ represent sliced indices. We encode the sliced tensors as 
\begin{align*}                          \Tilde{\boldsymbol{A}}^{(i)}_{-}&=\boldsymbol{A}^{(i)}_{1,-} + \boldsymbol{A}^{(i)}_{2,-}x \\ 
    \Tilde{\boldsymbol{B}}^{(j)}_{-}&=\boldsymbol{B}^{(j)}_{1,-} + \boldsymbol{B}^{(j)}_{2,-}x^{1+3} 
\end{align*}
where $i=1,2,3$ and $j=1,2,3,4$. $x$ can be substituted with a distinct evaluation point at each worker. After the local contraction operation, the result at each worker is: 
\begin{align*}
    \Tilde{\boldsymbol{\sigma}}&=\sum_{-}(\Tilde{\boldsymbol{A}}^{(1)}_{-}\Tilde{\boldsymbol{A}}^{(2)}_{-}\Tilde{\boldsymbol{A}}^{(3)}_{-}\Tilde{\boldsymbol{B}}^{(1)}_{-}\Tilde{\boldsymbol{B}}^{(2)}_{-}\Tilde{\boldsymbol{B}}^{(3)}_{-}\Tilde{\boldsymbol{B}}^{(4)}_{-})\boldsymbol{T}_{-} \\ 
    &= \boldsymbol{\sigma}_{1,1}+\boldsymbol{\sigma}_{2,1}x^3+\boldsymbol{\sigma}_{1,2}x^{4}+\boldsymbol{\sigma}_{2,2}x^{19}+ \ldots &= \text{poly}(x^{19}).
\end{align*}
Hence, with any $(19+1)$ successful workers with distinct evaluation points, we can perform polynomial interpolation to retrieve $\boldsymbol{\sigma}_{\text{final}}$. The gain compared to naive replication is
\begin{equation*}
    \Delta=4(f+1)-(f+20)=3f-16 \hspace{4pt}(\text{workers}),
\end{equation*}
which is consistent with Theorem~\ref{thm:hyperedge}.
\end{example}

\begin{remark}

Negative product term of $m_i$ in (12) of Theorem~\ref{thm:hyperedge} degrades the gain of the hyperedge code. Intuitively, that is because hyperedge code confines all desired terms each in different order terms hence resulting polynomial requires higher degree for larger $\{m_i\}$. On the contrary, 2-node code confines all desired terms in a single term of the resulting polynomial, therefore its gain (7) does not suffer such loss. However, we emphasize that our codes and their upper bounds of $f$-resilient number are not proven to be optimal nor tight.

\end{remark}

\subsection{Master Node Complexity}
Lastly, we show that encoding and decoding costs are negligible compared to computation-intensive tensor contraction. We will make an informal argument with as simple setting where all edges are $m$-node and $L$-dimensional, and all nodes have a degree $k$. 
The computation cost of tensor contraction is determined by the most expensive edge contraction~\cite{markovshi}. In our simplified setting, all edge contraction has the same cost. Contracting any index will result in an intermediate tensor with $m(k-1)$ open edges, as there are $m$ tensors with each $(k-1)$ edges not being contracted. The computational complexity for this is $O(L^{(k-1)m}\cdot L)$. On the other hand, encoding sliced tensors is equivalent to taking a linear combination of $L$ sliced tensors each of which has $k-1$ open edges. Hence, the cost of encoding is $O(L^{k-1}\cdot L)$, assuming $m$ is not exponentially large compared to $L$. Compared to contraction complexity of $O(L^{(k-1)m+1})$, encoding complexity of $O(L^k)$ is negligible. For example, for a typical PEPS tensor network with $k=4$ and $m=2$, contraction cost for an edge is $O(L^7)$ and encoding cost is $O(L^4)$. For decoding complexity, the final outcome of tensor network contraction in most quantum applications is a scalar or a vector, and thus decoding is minimal~\cite{polyinterpolation,polycode}.

\subsection{Existing matrix codes under our framework}
\begin{figure}[H]
    \centering
    \includegraphics[width=0.4\textwidth]{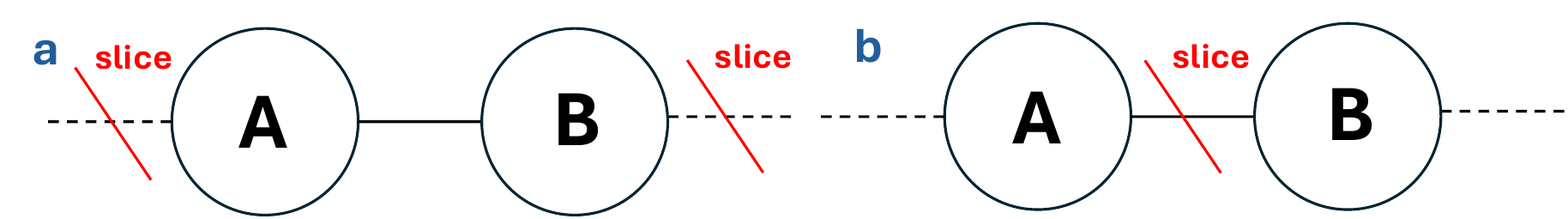}
    \caption{\small{Diagrammatic representation of ($\boldsymbol{a}$) polynomial code and ($\boldsymbol{b}$) MatDot code in tensor network formalism.}}
    \label{fig:matcodes}
    \vspace{-7pt}
\end{figure}
As the tensor network formalism is a general linear algebraic framework that also encompasses matrix multiplication, our work can be thought of as a generalized expansion of precedent matrix multiplication coding schemes.
For instance, in the tensor network formalism, MatDot codes\cite{matdot} slice a 2-node index between $\boldsymbol{A}$ and $\boldsymbol{B}$ (see Fig. \ref{fig:matcodes}), which corresponds to the case of $n=1$, $m_1 =2$, and $N =L_1$ for Theorem~\ref{thm:2-node}. In the formalism of our work, $f$-resilient number of the 2-node code is $f+ 2 L_1 -1$, which is consistent with the $f$-resilient number of MatDot codes~\cite{matdot}. 
On the other hand, polynomial code\cite{polycode} corresponds to slicing through open edges in tensor network formalism (see Fig. \ref{fig:matcodes}). This is a case we did not discuss in our work as open edges are not to be sliced as they do not reduce the complexity of the tensor network. Finally,
 $n$-matrix multiplication codes described in \cite{matdot} correspond to a chain in tensor network formalism and slicing through all edges (with alternating index dimension of $s$ and $t$). While this gave a rise to high $f$-resilient number in the original work\cite{matdot}, our work shines a new light on this problem that in practical algorithms, coding overhead can be much lower than $n$-matrix codes as we only slice through a select few edges instead of all edges~\cite{chenzhang}.

\section{Discussion and future works}
This work is the first attempt at applying coded computing 
for reliable tensor network contraction on large-scale parallel machines. 
There are many more exciting future directions to explore. While we provide a coding scheme that works only for non-adjacent edges with some practical justification, making it more general to include both adjacent and non-adjacent edge slicing would be an interesting problem. Another crucial research question is a lower bound on the $f$-resilient number of coded tensor network algorithms. While we provide two different achievable schemes in Theorem~\ref{thm:2-node} and Theorem~\ref{thm:hyperedge}, we do not know if the proposed approach is optimal or how far from optimality it is. Finally, as we are inspired by very practical applications, it would be valuable to implement the proposed coded computing strategies and perform an empirical evaluation  on real-world high-performance computing systems. 



\newpage

\bibliographystyle{IEEEtran}

\bibliography{IEEEabrv,references}

\appendices

\section{Proof of Theorem 1}\label{appendixA}

Let us refer to the sliced indices by number $i={1,2,\ldots,n}$. Each index $i$ is $L_i$ dimensional and shared by 2 tensors. For convenience, let us denote the tensors sharing index $1$ as $(\boldsymbol{A}^{(1)}_{-},\boldsymbol{A}^{(2)}_{-})$, the tensors sharing index $2$ as $(\boldsymbol{B}^{(1)}_{-},\boldsymbol{B}^{(2)}_{-})$, an the tensors sharing index $n$ as $(\boldsymbol{Z}^{(1)}_{-},\boldsymbol{Z}^{(2)}_{-})$. Each of the tensors is sliced by the selected index, for instance, $\boldsymbol{A}^{(1)}_{-}$ is sliced as $\boldsymbol{A}^{(1)}_{s_{1},-}$ for fixed value $s_1\in\{1,2,\ldots,L_1\}$ of index $1$. Then for fixed value $s_i$ for each index $i$, which span by $s_i\in\{1,2,\ldots,L_i\}$, equation for each sliced tensor network corresponding to the string $s_1s_2...s_n$ is
$$\boldsymbol{\sigma}_{s_{1}s_{2}...s_{n}}=\sum_{-}(\boldsymbol{A}^{(1)}_{s_{1},-}\boldsymbol{A}^{(2)}_{s_{1},-}\boldsymbol{B}^{(1)}_{s_{2},-}\boldsymbol{B}^{(2)}_{s_{2},-}...\boldsymbol{Z}^{(1)}_{s_{n},-}\boldsymbol{Z}^{(2)}_{s_{n},-})\boldsymbol{T}_{-},$$
where each index $i$ spans by $s_i\in\{1,2,\ldots,L_i\}$. Hence, there are $(N=\prod_{i=1}^{n}L_i)$ terms in the summation we need to retrieve.

For index $i$, we define two template polynomials:
$$
p_i^{(1)}(x)=1+x+x^2+\ldots+x^{L_i-1}
$$
$$
p_i^{(2)}(x)=x^{L_i-1}+x^{L_i-2}+x^{L_i-3}+\ldots+1.
$$

We encode for index $1$ with encoding polynomial $p_1^{(1)}(x)$ and $p_2^{(2)}(x)$ by putting each sliced tensors of $\boldsymbol{A}^{(1)}_{-}$ and $\boldsymbol{A}^{(2)}_{-}$ to the coefficients of $p_1^{(1)}(x)$ and $p_2^{(2)}(x)$ respectively as below:
$$
\Tilde{\boldsymbol{A}}^{(1)}_{-}=\boldsymbol{A}^{(1)}_{1,-}+x\boldsymbol{A}^{(1)}_{2,-}+\ldots+x^{L_1-1}\boldsymbol{A}^{(1)}_{L_1,-}=\sum_{k=1}^{L_1}x^{k-1}\boldsymbol{A}^{(1)}_{k,-}
$$
$$
\Tilde{\boldsymbol{A}}^{(2)}_{-}=x^{L_1-1}\boldsymbol{A}^{(2)}_{1,-}+x^{L_1-2}\boldsymbol{A}^{(2)}_{2,-}+..+\boldsymbol{A}^{(2)}_{L_1,-}=\sum_{k=1}^{L_1}x^{L_1-k}\boldsymbol{A}^{(2)}_{k,-}.
$$
Then, product of encoded tensors $\Tilde{\boldsymbol{A}}^{(1)}_{-} \Tilde{\boldsymbol{A}}^{(2)}_{-}$ is:
$$\Tilde{\boldsymbol{A}}^{(1)}_{-}\Tilde{\boldsymbol{A}}^{(2)}_{-}=(\sum_{k=1}^{L_1}x^{k-1}\boldsymbol{A}^{(1)}_{k,-})(\sum_{k=1}^{L_1}x^{L_1-k}\boldsymbol{A}^{(2)}_{k,-})
$$
$$
=x^{L_1-1}(\boldsymbol{A}^{(1)}_{1,-}\boldsymbol{A}^{(2)}_{1,-}+\boldsymbol{A}^{(1)}_{2,-}\boldsymbol{A}^{(2)}_{2,-}+\ldots+\boldsymbol{A}^{(1)}_{L_1,-}\boldsymbol{A}^{(2)}_{L_1,-})
$$
$$
+(\text{other order terms}) = poly(x^{2(L_1-1)}).
$$
All the desired terms of $\boldsymbol{A}^{(1)}_{s_1,-}\boldsymbol{A}^{(2)}_{s_1,-}(s_1\in\{1,2,...,L_1\})$ are confined in one coefficient of $(L_1-1)$ order term.

For index $2$, we use encoding polynomial $p^{(1)}_{2}(x^{L_1})$ and $p^{(2)}_{2}(x^{L_1})$:
$$
\Tilde{\boldsymbol{B}}^{(1)}_{-}=\boldsymbol{B}^{(1)}_{1,-}+\ldots+x^{L_1(L_2-1)}\boldsymbol{B}^{(1)}_{L_2,-}=\sum_{k=1}^{L_2}x^{L_1(k-1)}\boldsymbol{B}^{(1)}_{k,-}
$$
$$
\Tilde{\boldsymbol{B}}^{(2)}_{-}=x^{L_1(L_2-1)}\boldsymbol{B}^{(2)}_{1,-}+\ldots+\boldsymbol{B}^{(2)}_{L_2,-}=\sum_{k=1}^{L_2}x^{L_1(L_2-k)}\boldsymbol{B}^{(2)}_{k,-}.
$$
$$\Tilde{\boldsymbol{B}}^{(1)}_{-}\Tilde{\boldsymbol{B}}^{(2)}_{-}=(\sum_{k=1}^{L_2}x^{L_1(k-1)}\boldsymbol{B}^{(1)}_{k,-})(\sum_{k=1}^{L_2}x^{L_1(L_2-k)}\boldsymbol{B}^{(2)}_{k,-})
$$
$$
=x^{L_1(L_2-1)}(\boldsymbol{B}^{(1)}_{1,-}\boldsymbol{B}^{(2)}_{1,-}+\boldsymbol{B}^{(1)}_{2,-}\boldsymbol{B}^{(2)}_{2,-}+\ldots+\boldsymbol{B}^{(1)}_{L_2,-}\boldsymbol{B}^{(2)}_{L_2,-})
$$
$$
+(\text{other order terms}) = poly(x^{2L_1(L_2-1)}).
$$
Likewise, all the desired terms of $\boldsymbol{B}^{(1)}_{s_2,-}\boldsymbol{B}^{(2)}_{s_2,-}(s_2\in\{1,2,...,L_2\})$ are confined in one coefficient of $L_1(L_2-1)$ order term.

Furthermore, polynomial resulting from the product ($\Tilde{\boldsymbol{A}}^{(1)}_{-}\Tilde{\boldsymbol{A}}^{(2)}_{-}\Tilde{\boldsymbol{B}}^{(1)}_{-}\Tilde{\boldsymbol{B}}^{(2)}_{-}$) is:

$$
\Tilde{\boldsymbol{A}}^{(1)}_{-}\Tilde{\boldsymbol{A}}^{(2)}_{-}\Tilde{\boldsymbol{B}}^{(1)}_{-}\Tilde{\boldsymbol{B}}^{(2)}_{-}=x^{L_1L_2-1}(\sum_{s_1,s_2}\boldsymbol{A}^{(1)}_{s_1,-}\boldsymbol{A}^{(2)}_{s_1,-}\boldsymbol{B}^{(1)}_{s_2-}\boldsymbol{B}^{(2)}_{s_2,-})
$$
$$
+(\text{other order terms}) = poly(x^{2(L_1L_2-1)}).
$$
All desired terms of $(\boldsymbol{A}^{(1)}_{s_1,-}\boldsymbol{A}^{(2)}_{s_1,-}\boldsymbol{B}^{(1)}_{s_2-}\boldsymbol{B}^{(2)}_{s_2,-})$ for $s_1\in\{1,2,..,L_1\}$ $s_2\in\{1,2,..,L_2\}$ are confined in one coefficient of $(L_1L_2-1)$ order term.

For index $3$, encoding polynomials are $p^{(1)}_{3}(x^{L_1L_2})$ and $p^{(2)}_{3}(x^{L_1L_2})$. For index $i$, encoding polynomials are $p^{(1)}_{i}(x^{\prod_{j=1}^{i-1}L_j})$ and $p^{(2)}_{i}(x^{\prod_{j=1}^{i-1}L_j})$. For the last index $n$, encoded tensors are 
$$
\Tilde{\boldsymbol{Z}}^{(1)}_{-}=\sum_{k=1}^{L_n}x^{(k-1)\prod_{j=1}^{n-1}L_j}\boldsymbol{Z}^{(1)}_{k,-}
$$
$$
\Tilde{\boldsymbol{Z}}^{(2)}_{-}=\sum_{k=1}^{L_n}x^{(L_n-k)\prod_{j=1}^{n-1}L_j}\boldsymbol{Z}^{(2)}_{k,-}.
$$
And their product is
$$
\Tilde{\boldsymbol{Z}}^{(1)}_{-}\Tilde{\boldsymbol{Z}}^{(2)}_{-}=x^{(L_n-1)\prod_{j=1}^{n-1}L_j}(\boldsymbol{Z}^{(1)}_{1,-}\boldsymbol{Z}^{(2)}_{1,-}+\ldots+\boldsymbol{Z}^{(1)}_{L_n,-}\boldsymbol{Z}^{(2)}_{L_n,-}).
$$

As a result, an encoded tensor network produces a polynomial at the end of the contraction, which confines all desired outcome of sliced tensor networks in one coefficient of $((\prod_{i=1}^nL_i) - 1)$ order term.

$$\Tilde{\boldsymbol{\sigma}}=\sum_{-}(\Tilde{\boldsymbol{A}}^{(1)}_{-}\Tilde{\boldsymbol{A}}^{(2)}_{-}\Tilde{\boldsymbol{B}}^{(1)}_{-}\Tilde{\boldsymbol{B}}^{(2)}_{-}...\Tilde{\boldsymbol{Z}}^{(1)}_{-}\Tilde{\boldsymbol{Z}}^{(2)}_{-})\boldsymbol{T}_{-}$$
$$
=\sum_{-}[(\sum_{s_1,s_2,..,s_n}\boldsymbol{A}^{(1)}_{s_1,-}\boldsymbol{A}^{(2)}_{s_1,-}\boldsymbol{B}^{(1)}_{s_2,-}..\boldsymbol{Z}^{(2)}_{s_n,-})x^{(\prod_{i=1}^nL_i)-1}+..]\boldsymbol{T}_{-}
$$
$$
=(\sum_{s_1,s_2,..,s_n}\boldsymbol{\sigma}_{s_{1}s_{2}...s_{n}})x^{(\prod_{i=1}^nL_i)-1}+(\text{other order terms})
$$
$$
=\boldsymbol{\sigma}_{\text{final}}x^{(\prod_{i=1}^nL_i)-1}+(\text{other order terms})=poly(x^{2(\prod_{i=1}^nL_i)-2}).
$$
Hence, with any successful $(2(\prod_{i=1}^nL_i)-2+1)$ encoded tensor networks each using different arbitrary constant $x$, coefficient of $((\prod_{i=1}^nL_i) - 1)$ order term in the resulting polynomial can be retrieved, which is the final outcome of the original tensor network.

Therefore, for $f$ node failures, $f$-resilient number is
$$
f+2\prod_{i=1}^nL_i-1
$$
for 2-node code.

\section{Proof of Theorem 2}\label{appendixB}

Let us refer to the sliced indices by number $i={1,2,\ldots,n}$. Each index $i$ is $L_i$ dimensional and shared by $m_i$ tensors. For convenience, let us denote the tensors sharing index $1$ as $(\boldsymbol{A}^{(1)}_{-},\boldsymbol{A}^{(2)}_{-},\ldots,\boldsymbol{A}^{(m_1)}_{-})$, and tensors sharing index $2$ as $(\boldsymbol{B}^{(1)}_{-},\boldsymbol{B}^{(2)}_{-},\ldots,\boldsymbol{B}^{(m_2)}_{-})$. We will denote tensors sharing index $n$ as $(\boldsymbol{Z}^{(1)}_{-},\boldsymbol{Z}^{(2)}_{-},\ldots,\boldsymbol{Z}^{(m_n)}_{-})$. Each of the tensors is sliced by the selected index, for instance, $\boldsymbol{A}^{(1)}_{-}$ is sliced as $\boldsymbol{A}^{(1)}_{s_{1},-}$ for fixed value $s_1\in\{1,2,\ldots,L_1\}$ of index $1$. Then, for fixed value $s_i$ for each index $i$, the equation for each sliced tensor network corresponding to the string $s_1s_2...s_n$ is
$$\boldsymbol{\sigma}_{s_{1}s_{2}...s_{n}}=\sum_{-}(\prod_{j=1}^{m_1}\boldsymbol{A}^{(j)}_{s_{1},-}\prod_{k=1}^{m_2}\boldsymbol{B}^{(k)}_{s_{2},-}...\prod_{l=1}^{m_n}\boldsymbol{Z}^{(l)}_{s_{n},-})\boldsymbol{T}_{-},$$
where each index $i$ spans by $s_i\in\{1,2,\ldots,L_i\}$. Hence, there are $(N=\prod_{i=1}^{n}L_i)$ product terms in the summation we need to retrieve.

For an index $i$, we define a template polynomial $p_{i}(x)$ to encode tensors sharing the corresponding index as 
$$
p_{i}(x)=1+x+x^{1+m_i}+x^{1+m_i+m_i^2}+\ldots+x^{1+m_i+\dots+m_i^{L_i-2}}$$
$$=\sum_{j=1}^{L_i}x^{\frac{m_i^{j-1}-1}{m_i-1}}.
$$
For instance, encoding polynomial for index $1$ is $p_1(x)$. That is, we encode each tensor sliced by the index $1$ by putting the sliced tensors on each of the coefficients of encoding polynomial $p_{1}(x)$:
$$
\Tilde{\boldsymbol{A}}^{(k)}_{-}=\sum_{j=1}^{L_1}\boldsymbol{A}^{(k)}_{j,-}x^{\frac{m_1^{j-1}-1}{m_1-1}}(k=1,2,..,m_1)
$$
$$
=\boldsymbol{A}^{(k)}_{1,-}+\boldsymbol{A}^{(k)}_{2,-}x+\boldsymbol{A}^{(k)}_{3,-}x^{1+m_i}+\ldots+\boldsymbol{A}^{(k)}_{L_1,-}x^{\frac{m_1^{L_1-1}-1}{m_1-1}},
$$
for $k\in\{1,2,...,m_1\}$.

Then, the product of the encoded tensors results in a polynomial as below:
$$
\prod_{k=1}^{m_1}\Tilde{\boldsymbol{A}}^{(k)}=\prod_{k=1}^{m_1}\boldsymbol{A}^{(k)}_{1,-}+x^{m_1}\prod_{k=1}^{m_1}\boldsymbol{A}^{(k)}_{2,-}+x^{m_1^2+m_1}\prod_{k=1}^{m_1}\boldsymbol{A}^{(k)}_{3,-}
$$
$$
+x^{m^3_1+m^2_1+m_1}\prod_{k=1}^{m_1}\boldsymbol{A}^{(k)}_{4,-}+\ldots+x^{m_1^{L_1-1}+...+m_1}\prod_{k=1}^{m_1}\boldsymbol{A}^{(k)}_{L_1,-}
$$
$$
+(\text{other order terms}) = poly(x^{m_1^{L_1-1}+...+m_1}=x^{\frac{m_1^{L_1}-m_1}{m_1-1}}).
$$
Each term of $\prod_{k=1}^{m_1}\boldsymbol{A}^{(k)}_{s_1,-}$ for $s_1\in\{1,2,\ldots,L_1\}$ is confined on different order term only by itself. That is because order of individual term of the encoding polynomial $p_1(x)$ is larger than $\times m_1$ of previous order.

For index $2$, we use encoding polynomial $p_2(x^{\frac{m_1^{L_1}-1}{m_1-1}})$ where
$$
p_{2}(x^{\frac{m_1^{L_1}-1}{m_1-1}})=1+x^{\frac{m_1^{L_1}-1}{m_1-1}}+x^{(1+m_2)\frac{m_1^{L_1}-1}{m_1-1}}+x^{(1+m_2+m_2^2)\frac{m_1^{L_1}-1}{m_1-1}}
$$
$$
+\ldots+x^{(1+m_2+\dots+m_2^{L_2-2})\frac{m_1^{L_1}-1}{m_1-1}}=\sum_{j=1}^{L_2}x^{\frac{m_1^{L_1}-1}{m_1-1}\frac{m_2^{j-1}-1}{m_2-1}}.
$$
Same as for index $1$, we encode each tensor of index $2$ by putting the sliced tensors on each of the coefficients of encoding polynomial $p_2(x^{\frac{m_1^{L_1}-1}{m_1-1}})$:
$$
\Tilde{\boldsymbol{B}}^{(k)}_{-}=\sum_{j=1}^{L_2}\boldsymbol{B}^{(k)}_{j,-}x^{\frac{m_1^{L_1}-1}{m_1-1}\frac{m_2^{j-1}-1}{m_2-1}}(k=1,2,..,m_2)
$$
$$
=\boldsymbol{B}^{(k)}_{1,-}+\boldsymbol{B}^{(k)}_{2,-}x^{\frac{m_1^{L_1}-1}{m_1-1}}+\boldsymbol{B}^{(k)}_{3,-}x^{(1+m_2)\frac{m_1^{L_1}-1}{m_1-1}}$$
$$
+\boldsymbol{B}^{(k)}_{4,-}x^{(1+m_2+m^2_2)\frac{m_1^{L_1}-1}{m_1-1}}+\ldots+\boldsymbol{B}^{(k)}_{L_2,-}x^{\frac{m_1^{L_1}-1}{m_1-1}\frac{m_2^{L_2-1}-1}{m_2-1}}.
$$
Then the product of the encoded tensors for index $2$ results in a polynomial as below:
$$
\prod_{k=1}^{m_2}\Tilde{\boldsymbol{B}}^{(k)}=\prod_{k=1}^{m_2}\boldsymbol{B}^{(k)}_{1,-}+x^{m_2\frac{m_1^{L_1}-1}{m_1-1}}\prod_{k=1}^{m_1}\boldsymbol{B}^{(k)}_{2,-}
$$
$$+x^{(m_2^2+m_2)\frac{m_1^{L_1}-1}{m_1-1}}\prod_{k=1}^{m_1}\boldsymbol{B}^{(k)}_{3,-}
+x^{(m^3_2+m^2_2+m_2)\frac{m_1^{L_1}-1}{m_1-1}}\prod_{k=1}^{m_1}\boldsymbol{B}^{(k)}_{4,-}
$$
$$+\ldots+x^{(m_2^{L_2-1}+...+m_2)\frac{m_1^{L_1}-1}{m_1-1}}\prod_{k=1}^{m_1}\boldsymbol{B}^{(k)}_{L_2,-}+(\text{other order terms})
$$
$$
= poly(x^{\frac{m_1^{L_1}-1}{m_1-1}\frac{m_2^{L_2}-m_2}{m_2-1}}).
$$
Just as the case of index $1$, each term of $\prod_{k=1}^{m_2}\boldsymbol{B}^{(k)}_{s_2,-}$ for $s_2\in\{1,2,\ldots,L_2\}$ is confined on different order term only by itself. Likewise, this is because order of individual term of the encoding polynomial $p_2(x^{\frac{m_1^{L_1}-1}{m_1-1}})$ is larger than $\times m_2$ of previous order.

Then, polynomial resulting from $(\prod_{k=1}^{m_1}\Tilde{\boldsymbol{A}}^{(k)}\prod_{k=1}^{m_2}\Tilde{\boldsymbol{B}}^{(k)})
$ confines every individual term of $(\prod_{j=1}^{m_1}\boldsymbol{A}^{(j)}_{s_{1},-}\prod_{k=1}^{m_2}\boldsymbol{B}^{(k)}_{s_{2},-})$ for $s_1\in\{1,2,\ldots,L_1\}, s_2\in\{1,2,\ldots,L_2\}$ each in different order term. That is because order of individual terms of $p_2(x^{\frac{m_1^{L_1}-1}{m_1-1}})$ is larger than $[m_2\times(\text{previous order})+\frac{m_1^{L_1}-m_1}{m_1-1}]$.

To generalize, tensors sliced by index $i$ are encoded by encoding polynomial of 
$$
p_i(x^{\prod_{j=1}^{i-1}\frac{m_j^{L_j}-1}{m_j-1}}),
$$
by putting the sliced tensors on each of the coefficients of the encoding polynomial. For instance, encoded tensors for last index $n$ are as follows:
$$
\Tilde{\boldsymbol{Z}}^{(k)}_{-}=\sum_{j=1}^{L_n}\boldsymbol{Z}^{(k)}_{j,-}x^{\prod_{j=1}^{n-1}\frac{m_j^{L_j}-1}{m_j-1}} (k=1,2,..,m_n)
$$

As a result, an encoded tensor network results as a polynomial at the end of the contraction, which confines every desired outcome of sliced tensor networks each in different order term as below: $$\Tilde{\boldsymbol{\sigma}}=\sum_{-}(\prod_{j=1}^{m_1}\Tilde{\boldsymbol{A}}^{(j)}_{s_{1},-}\prod_{k=1}^{m_2}\Tilde{\boldsymbol{B}}^{(k)}_{s_{2},-}...\prod_{l=1}^{m_n}\Tilde{\boldsymbol{Z}}^{(l)}_{s_{n},-})\boldsymbol{T}_{-}$$
$$=\sum_{-}[\prod_{j=1}^{m_1}\boldsymbol{A}^{(j)}_{1,-}\prod_{k=1}^{m_2}\boldsymbol{B}^{(k)}_{1,-}...\prod_{l=1}^{m_n}\boldsymbol{Z}^{(l)}_{1,-}+\ldots
$$
$$
+x^{(\prod_{j=1}^{n}\frac{m_j^{L_j}-1}{m_j-1})-1}\boldsymbol{A}^{(j)}_{L_1,-}\prod_{k=1}^{m_2}\boldsymbol{B}^{(k)}_{L_2,-}...\prod_{l=1}^{m_n}\boldsymbol{Z}^{(l)}_{L_n,-}]\boldsymbol{T}_{-}
$$
$$
=\boldsymbol{\sigma}_{11...1}+\ldots+\boldsymbol{\sigma}_{L_1L_2...L_n}x^{(\prod_{j=1}^{n}\frac{m_j^{L_j}-1}{m_j-1})-1}
$$
$$
=poly(x^{(\prod_{j=1}^{n}\frac{m_j^{L_j}-1}{m_j-1})-1})
$$
Every coefficient of this resulting polynomial can be retrieved by polynomial interpolation with $(\prod_{i=1}^{n}\frac{m_i^{L_i}-1}{m_i-1})$ different arbitrary constants. Hence, with any successful $(\prod_{i=1}^{n}\frac{m_i^{L_i}-1}{m_i-1})$ encoded tensor networks each using different arbitrary constant $x$, all the desired outcomes of sliced tensor networks can be retrieved and the final outcome of original tensor network can be acquired by their summation.

Therefore, for $f$ node failures, $f$-resilient number is
$$
f+\prod_{i=1}^{n}\frac{m_i^{L_i}-1}{m_i-1}.
$$
for hyperedge code.

\end{document}